%% file: main.tex
\documentclass[10pt]{article} 
\usepackage[preprint]{tmlr}

\input{math_commands.tex}

\usepackage{hyperref}
\hypersetup{
	colorlinks=true,
        breaklinks=true,
        citecolor={black!10!blue},
}
\usepackage{url}

\newcommand{\on}[1]{\operatorname{#1}}

\title{Fully Decentralized Cooperative Multi-Agent Reinforcement Learning: A Survey}

\author{\name Jiechuan Jiang$^\dagger$ \email jiechuan.jiang@pku.edu.cn \\
      \addr School of Computer Science\\
      Peking University
      \AND
      \name Kefan Su$^\dagger$ \email sukefan@pku.edu.cn \\
      \addr School of Computer Science\\
      Peking University
      \AND
      \name Zongqing Lu \email zongqing.lu@pku.edu.cn\\
      \addr School of Computer Science\\
      Peking University
      \AND
      \addr $^\dagger$\emph{Equal contribution, alphabetic order}
      }

\def\month{MM}  
\def\year{YYYY} 
\def\openreview{\url{https://openreview.net/forum?id=XXXX}} 

\begin{document}

\maketitle

\begin{abstract}

Cooperative multi-agent reinforcement learning is a powerful tool to solve many real-world cooperative tasks, but restrictions of real-world applications may require training the agents in a fully decentralized manner. Due to the lack of information about other agents, it is challenging to derive algorithms that can converge to the optimal joint policy in a fully decentralized setting. Thus, this research area has not been thoroughly studied. In this paper, we seek to systematically review the fully decentralized methods in two settings: maximizing a shared reward of all agents and maximizing the sum of individual rewards of all agents, and discuss open questions and future research directions.

\end{abstract}

\section{Introduction}

Many real-world applications require that multiple agents cooperatively accomplish a task, including traffic signal control \citep{xu2021hierarchically}, power dispatch \citep{wang2021multi}, finance \citep{fang2023learning}, and robot control \citep{orr2023multi}. Recently, accompanied by the maturity of deep learning techniques, multi-agent reinforcement learning (MARL) has been widely applied to such cooperative tasks, where a group of agents interacts with a common environment, each agent decides its local action, and they are trained to maximize a shared reward or the sum of individual rewards. 

There are two main paradigms of cooperative MARL: centralized training with decentralized execution (CTDE) and fully decentralized learning, according to whether the information of other agents, \textit{e.g.}, the actions of other agents, can be obtained during the training process. In CTDE methods, each agent has access to global information during the training but only relies on its local information to make decisions during execution. A lot of CTDE methods have been proposed \citep{lowe2017multi,sunehag2018value, rashid2018qmix, son2019qtran, iqbal2019actor, wang2020qplex, rashid2020weighted, wang2020dop, zhang2021fop, su2021divergence, peng2021facmac,li2022difference,wang2023more, wang2023mutual} and achieve significant performance in multi-agent benchmarks, \textit{e.g.}, StarCraft Multi-Agent Challenge \citep{samvelyan19smac} and Google research football \citep{kurach2020google}. However, in some scenarios where global information is unavailable or the number of agents is dynamically changing, the centralized modules lose effectiveness, and fully decentralized learning is necessary \citep{zhang2021multi}. In fully decentralized learning, each agent cannot obtain any information from other agents in both training and execution. Other agents have to be treated as a part of the environment but are updating their policies during the training. Thus the environment becomes non-stationary from the perspective of individual agents \citep{foerster2017stabilising, jiangi2q}, which violates the assumptions of almost all existing reinforcement learning methods and makes it challenging to derive algorithms that can converge to the optimal joint policies in the fully decentralized setting. Perhaps due to this reason, research on decentralized learning algorithms is limited. Therefore, in this paper, we provide an overview of cooperative MARL, with a focus on fully decentralized learning algorithms, hoping to assist researchers in gaining a clear understanding and generating more interest in this challenging yet meaningful research direction.

Fully decentralized cooperative MARL is commonly formulated into two settings: maximizing a shared reward of all agents (called shared reward setting) and maximizing the sum of individual rewards of all agents (called reward sum setting). The shared reward setting is the most popular formulation in the recent MARL methods with deep learning. We review the work of shared reward setting within two representative frameworks: value-based methods and policy-based methods. Value-based methods focus on mitigating the impact of non-stationary transition probabilities. Policy-based methods investigate how to guarantee the monotonic improvement of the joint policies in fully decentralized optimization. Most of the methods in these two categories propose new value iterations or new optimization objectives of policies, thus they can be naturally combined with deep neural networks and are practical in high-dimensional complex tasks. Then we present the work in the reward sum setting, where the target of algorithms is to find a Nash equilibrium for all agents. We also categorize existing studies into value-based algorithms and policy-based algorithms. The value-based algorithms use Q-learning \citep{watkins1992q} to obtain the best response policy of the other agents' policies. The policy-based algorithms control the difference between the current policy and the best response by the gradient dominance condition. 

We structure the paper as follows. Section \ref{sec:background} covers the background on single-agent RL and cooperative multi-agent RL, highlighting the fully decentralized formulation. In Section \ref{sec:shared} and Section \ref{sec:sum}, we respectively review the methods of shared reward setting and reward sum setting. Finally, in Section \ref{sec:discuss} we discuss limitations, open questions, and future research directions for fully decentralized MARL.

\section{Background}
\label{sec:background}

\subsection{Single-Agent RL}

Reinforcement learning is usually formulated as a Markov decision process, which is defined as a tuple $\left\langle\mathcal{S}, \mathcal{A}, P, r, \gamma, \rho_0\right\rangle$. $\mathcal{S}$ and $\mathcal{A}$ respectively denote the state and action spaces. $P(s'|s,a)$ denotes the transition probability from the state $s \in \mathcal{S}$ to the next state $s' \in \mathcal{S}$ for the given action $a \in \mathcal{A}$. $r(s,a,s')$ is the reward function for evaluating transitions. $\gamma \in [0,1)$ is the discount factor, and $\rho_0$ is the distribution of initial states. At each timestep $t$, the agent receives the state $s_t$ and selects an action $a_t$ according to its policy $\pi$. The environment transitions to the next state $s_{t+1}$ according to transition probability $P(s'|s,a)$, and the agent receives a reward $r(s_t,a_t,s_{t+1})$. Reinforcement learning aim at learning a policy $\pi$ to maximize the expected discounted return
$$J(\pi)=\mathbb{E}_{\pi}\left[\sum_t^{\infty} \gamma^t r\left(s_t, a_t, s_{t+1}\right)\right].$$

The action-value function $Q^{\pi}$ is defined as the expected return if the agent starts from state $s$, takes action $a$, and then forever acts according to policy $\pi$:
$$Q^\pi(s, a)=\mathbb{E}_{\pi}\left[\sum_{t=0}^{\infty} \gamma^t r\left(s_t, a_t, s_{t+1}\right) \mid s_0=s, a_0=a\right].$$
The state-value function $V^{\pi}$ is defined as the expected return if the agent starts from state $s$ and always acts according to policy $\pi$:
$$V^\pi(s)=\mathbb{E}_{a \sim \pi}\left[Q^\pi(s, a)\right]=\mathbb{E}_{\pi}\left[\sum_{t=0}^{\infty} \gamma^t r\left(s_t, a_t, s_{t+1}\right) \mid s_0=s\right].$$
The advantage function $A^{\pi}$ describes how much better it is to take an action $a$ over acting according to $\pi$:
$$A^\pi(s, a)=Q^\pi(s, a)-V^\pi(s).$$
The functions corresponding to the optimal policy $\pi^*$ are defined as the optimal Q-function $Q^*$ and the optimal V-function $V^*$.

RL algorithms can be commonly categorized into two frameworks, value-based and policy-based methods, according to whether a policy is explicitly learned. We introduce the most typical methods for both types. Q-learning \citep{watkins1992q} is a representative value-based method. It updates the optimal Q-function $Q^*$ using the Bellman operator:
$$\mathcal{T} Q^*(s, a)=\mathbb{E}_{s^{\prime} \sim P}\left[r+\gamma \max _{a^{\prime}} Q^*\left(s^{\prime}, a^{\prime}\right)\right].$$
Under this Bellman iteration, Q-learning is proved to converge to the optimal Q-function $Q^*$ with finite state and action spaces. The optimal policy in Q-learning is deterministic and can be derived by greedily selecting the action with the highest Q-value $\pi(s) = \max_{a} Q^*\left(s, a\right)$. Q-learning is an off-policy algorithm, which can learn using the experiences collected by any policy. 

Policy-based methods directly learn a policy $\pi_{\theta}$ parameterized by $\theta$ using policy gradient. The most straightforward method is REINFORCE \citep{williams1992simple}, and with the advantage function the policy gradient $\nabla J(\theta)$ can be further formulated as 
$$\nabla J(\theta)=\mathbb{E}_{a \sim \pi_\theta(\cdot \mid s)}\left[A^{\pi_\theta}(s, a) \nabla \log \pi_\theta(a \mid s)\right].$$ 
However, as the policy is parameterized by $\theta$, the update of $\theta$ according to the policy gradient may dramatically change the policy. Thus policy gradient methods are hard to guarantee monotonic policy improvement.
TRPO \citep{TRPO} updates the policy by taking the largest step possible to improve performance with the constraint of KL-divergence between new and old policies
$$\theta_{k+1}=\arg \max _\theta \mathbb{E}_{s, a \sim \pi_{\theta_k}}\left[\frac{\pi_\theta(a \mid s)}{\pi_{\theta_k}(a \mid s)} A^{\pi_{\theta_k}}(s, a)\right], \quad s.t. \quad  \mathbb{E}_{s \sim \pi_{\theta_k}}\left[D_{\operatorname{KL}}\left(\pi_\theta(\cdot \mid s) \| \pi_{\theta_k}(\cdot \mid s)\right)\right] \leq \delta,$$
where $\pi_{\theta_k}$ is the old policy for experience collection. TRPO can guarantee monotonic improvement but has poor computation efficiency due to second-order optimization. PPO \citep{PPO} is a simple and empirical approximation of TRPO, which uses a clipping trick in the objective function to make sure the new policy is close to the old policy. The objective can be written as
$$\mathcal{L}(\pi_\theta)=\mathbb{E}_{s,a\sim \pi_{\theta_k}} \left[ \min \left(\frac{\pi_\theta(a \mid s)}{\pi_{\theta_k}(a \mid s)} A^{\pi_{\theta_k}}(s, a), \quad \operatorname{clip}\left(\frac{\pi_\theta(a \mid s)}{\pi_{\theta_k}(a \mid s)}, 1-\epsilon, 1+\epsilon\right) A^{\pi_{\theta_k}}(s, a)\right) \right],$$
where $\epsilon$ controls how far away $\pi_{\theta}$ is allowed to deviate from $\pi_{\theta_k}$. Policy-based methods are usually on-policy.

\subsection{Multi-Agent RL}

Extending single-agent RL to multi-agent RL, we consider multi-agent MDP $\left\langle\mathcal{S}, \mathcal{A}, P, \boldsymbol{r}, \gamma, \rho_0, N \right\rangle$. $N$ is the agent number. $\mathcal{S}$ denotes the state space. $\mathcal{A}:=\mathcal{A}_1 \times \cdots \times \mathcal{A}_N$ denotes the joint action space, and $\mathcal{A}_i$ is the action space of agent $i$. Each agent $i$ decides its own action $a_i \in \mathcal{A}_i$ according to its policy $\pi_i$. $P(s'|s,\boldsymbol{a})$ denotes the transition probability from the state $s \in \mathcal{S}$ to the next state $s' \in \mathcal{S}$ for the joint action $\boldsymbol{a}$. $\boldsymbol{r}(s,\boldsymbol{a},s') = [r_1,\cdots,r_i,\cdots,r_N]$ is the rewards of all agents. $\gamma \in [0,1)$ is the discount factor, and $\rho_0$ is the distribution of initial states. Cooperative MARL learns the joint policy $\boldsymbol{\pi}$ to maximize the expected discounted return of the sum of agents' rewards
$$J(\boldsymbol{\pi})=\mathbb{E}_{\boldsymbol{\pi}}\left[\sum_t^{\infty} \gamma^t \sum_{i = 1}^{N} r_{i}(s_t,\boldsymbol{a}_t,s_{t+1})\right].$$
If the agents' rewards are always the same, \textit{i.e.}, $r_1=r_2=,\cdots,=r_N$, the formulation is referred to as the shared reward setting, where all agents maximize the global objective, which is the most popular setting of MARL methods. Without this restriction, the formulation is called the reward sum setting. The former is a special case of the latter.

According to whether the information of other agents can be obtained during the training process, cooperative MARL algorithms can be divided into two categories: centralized training with decentralized execution (CTDE) and fully decentralized learning. In CTDE methods, each agent has access to global information during the training and only relies on its local information for decision-making in the execution. The most popular framework in CTDE is value decomposition \citep{sunehag2018value,rashid2018qmix,son2019qtran,wang2020qplex,rashid2020weighted,yang2020qatten}, where the joint Q-function is factorized into individual Q-functions by a mixer
$$Q_{tot}(s, \boldsymbol{a})=\operatorname{mixer}\left(Q_1\left(s, a_1\right), Q_2\left(s, a_2\right), \cdots, Q_N\left(s, a_N\right)\right).$$
The mixer should ensure that an $\argmax$ operation performed on $Q_{tot}$ yields the same joint action as a set of individual $\argmax$ operations performed on each $Q_i$. The mixer can be a sum operation \citep{sunehag2018value}, a weighted sum operation with positive weights produced by attention-network \citep{yang2020qatten}, or a neural network with positive weights produced by hyper-network \citep{rashid2018qmix}. Multi-agent actor-critic methods adopt vanilla centralized critic \citep{lowe2017multi,foerster2018counterfactual,iqbal2019actor} or value-decomposition critic \citep{wang2020dop,peng2021facmac}. FOP \citep{zhang2021fop} independently decomposes the joint policy into individual policies, and MACPF \citep{wang2023more} considers the dependency between individual policies in the decomposition. MAPPO \citep{yu2022surprising} extends PPO to the multi-agent setting by the centralized state-value function, and HAPPO \cite{kuba2021trust} decomposes the joint advantage function into individual advantage factions and sequentially updates the individual policies. Some methods \citep{zhang2018fully,konan2021iterated,li2020multi} require information sharing with neighboring agents according to a time-varying communication channel in both training and execution, which are categorized as networked agent setting and beyond the scope of decentralized execution.

\textbf{In fully decentralized learning, each agent cannot obtain any information from other agents in both training and execution and independently updates its own policy to maximize the sum of all agents' rewards.} Each agent is not allowed to share its actions, experiences, neural network parameters, etc, with other agents, and has to treat other agents as a part of the environment. Since all agents are updating their policies during the training, the environment becomes non-stationary from the individual agent's perspective, making it hard to develop algorithms that can converge to the optimal joint policy in a fully decentralized way. In the next two sections, we respectively review the existing algorithms that try to tackle this problem in the \textit{shared reward} setting and \textit{reward sum} setting.

\section{Shared Reward Setting}
\label{sec:shared}


\subsection{Value-Based Methods}

In fully decentralized learning, the transition probability from the perspective of each agent $i$ is
$$ P_{i}\left(s^{\prime} | s, a_i\right)={\sum}_{\boldsymbol{a}_{-i}}P\left(s^{\prime} | s, a_i,{a}_{-i}\right)  {\pi}_{-i}({a}_{-i} | s),$$
where ${a}_{-i}$ and ${\pi}_{-i}$ respectively denote the joint action and joint policy of all agents except agent $i$. $P_{i}$ depends on the policies of other agents ${\pi}_{-i}$. As other agents are updating their policies continuously, $P_i$ becomes non-stationary. Under the non-stationary transition probabilities, if each agent $i$ performs independent Q-learning (IQL) \citep{tan1993multi}
$$\mathcal{T} Q_i(s,a_i) = \mathbb{E}_{P_i(s'|s,a_i)}\left [r + \gamma \underset{a'_{i}}{\max}Q_i(s',a'_{i}) \right ],$$
the convergence of Q-function is not guaranteed. Q-learning is off-policy and commonly learns from experiences stored in a replay buffer. However, the experiences in the replay buffer are collected under changing transition probabilities $P_{i}$, thus the transition probabilities in the replay buffer are not only non-stationary but also obsolete.
Fingerprints \citep{foerster2017stabilising} adds iteration number and exploration rate to the state as conditions to disambiguate the age of experiences to alleviate the problem of obsolescence. Lenient IQL \citep{palmer2018lenient} uses a decayed leniency value to motivate the agents to focus on fresh experiences. MA2QL \citep{su2022ma2ql} lets the agents alternately perform IQL. While one agent is updating its Q-function, other agents keep their Q-functions unchanged. MA2QL guarantees the convergence to a Nash equilibrium, but the converged equilibrium may not be the optimal one when there are multiple Nash equilibria. Moreover, to obtain the theoretical guarantee, it has to be trained in an on-policy manner and each agent should collect its replay buffer from scratch at each training turn, which leads to poor sample efficiency. Therefore, to address this issue, MA2QL is trained in an off-policy manner when combined with neural networks. Distributed IQL \citep{lauer2000algorithm} is a new operator
$$\mathcal{T} Q_i(s,a_i) = \max\left(Q_i(s,a_i), \,r + \gamma \underset{a'_{i}}{\max}Q_i(s',a'_{i}) \right),$$
which can guarantee to converge to the optimal joint policy in deterministic environments. However, Distributed IQL fails in stochastic environments. Hysteretic IQL \citep{matignon2007hysteretic} is an extension of Distributed IQL, which applies different learning rates $w(s,a_i)$ to different experiences:
\begin{align*}
w(s, a_i) = \left\{\begin{matrix}
1 & \text{if }  r + \gamma \underset{a'_{i}}{\max}Q_i(s',a'_{i})  > Q_i\left(s, a_i\right)\\ 
\lambda < 1 & \text{else}.
\end{matrix}\right.
\end{align*}
If $\lambda = 0$, Hysteretic IQL degenerates to Distributed IQL. Hysteretic IQL mitigates the overestimation of Distributed IQL and is more robust in stochastic environments. I2Q \citep{jiangi2q} lets each agent perform IQL on ideal transition probabilities, which are defined as 
$$P\left(s^{\prime} | s, a_i, {\pi}^*_{-i}(s,a_i)\right), \quad {\pi}_{-i}^*(s,a_i) = \underset{{a}_{-i}}{\arg \max}Q^*(s,a_i,{a}_{-i}).$$
I2Q is proven to converge to the optimal joint policy under ideal transition probabilities. I2Q provides a method to obtain the ideal transition probabilities in deterministic environments by learning a value function $Q^{\mathrm{ss}}_i\left(s, s^{\prime}\right)$ using the following operator
$$\mathcal{T} Q^{\mathrm{ss}}_i\left(s, s^{\prime}\right)=r+\gamma \max _{s^{\prime \prime} \in \mathcal{N}\left(s^{\prime}\right)} Q^{\mathrm{ss}}_i\left(s^{\prime}, s^{\prime \prime}\right),$$
where $\mathcal{N}$ is the neighboring state set. In deterministic environments, under ideal transition probabilities, the state transitions to
$$s'^{*} = \underset{s' \in \mathcal{N}(s,a_i)}{\arg \max} Q^{\mathrm{ss}}_i(s,s').$$
However, how to obtain ideal transition probabilities in stochastic environments is a remaining question. 
For stochastic environments, BQL \citep{jiang2023best} proposes a new operator
$$\mathcal{T} Q_i(s,a_i) = \max\left(Q_i(s,a_i),\,\mathbb{E}_{\tilde{P}_i(s'|s,a_i)}\left [r + \gamma \underset{a'_{i}}{\max}Q_i(s',a'_{i}) \right ] \right),$$
where $\tilde{P}_i$ is one randomly selected of possible transition probabilities. BQL can converge to the optimal joint policy in both deterministic and stochastic environments if all possible transition probabilities can be sampled. Distributed IQL is a special case of BQL when the environment is deterministic. However, in neural network implementation, due to sample efficiency, BQL does not maintain multiple replay buffers to represent different transition probabilities but maintains only one buffer like IQL. The non-stationarity in the replay helps BQL sample different possible transition probabilities.

\subsection{Policy-Based Methods}

In the shared reward setting, policy-based methods are usually extended from single-agent RL. Independent learning is a straight but effective idea for fully decentralized learning. Recently, independent PPO (IPPO) \citep{IPPO} has attracted the attention of the MARL community. The algorithm of IPPO is that each agent $i$ updates its policy $\pi_i$ with PPO \citep{PPO}:
\begin{align}
    \mathcal{L}_i(\pi_i) = \mathbb{E}_{s,a_i} \left[ \min \left( \frac{\pi_i(a_i|s)}{\pi^{\operatorname{old}}_i(a_i|s)} A^{\operatorname{old}}_i(s,a_i) , \operatorname{clip} \left( \frac{\pi_i(a_i|s)}{\pi^{\operatorname{old}}_i(a_i|s)}, 1-\epsilon, 1+ \epsilon \right) A^{\operatorname{old}}_i(s,a_i)\right) \right]. \notag
\end{align}
IPPO obtains good performance, comparable to CTDE methods, in the popular MARL benchmark SMAC \citep{SMAC} with such a simple implementation. However, IPPO is still a heuristic algorithm troubled by the non-stationary problem.

DPO \citep{DPO} tries to solve the non-stationary problem following the idea of using a surrogate function as in TRPO \citep{TRPO} for single-agent RL. Suppose we use TRPO to learn a joint policy in a centralized manner, then we have the objective
\begin{align}
   &  J(\bm{\pi}) - J(\bm{\pi}^{\operatorname{old}}) \ge \mathcal{L}^{\operatorname{joint}}_{\bm{\pi}^{\operatorname{old}}}(\bm{\pi}) - C \cdot D_{\operatorname{KL}}^{\operatorname{max}}(\bm{\pi}^{\operatorname{old}} \| \bm{\pi}) = S^{\operatorname{TRPO}}(\bm{\pi}, \bm{\pi}^{\operatorname{old}}) \notag \\
    & {\text{where} \,\,}
     \mathcal{L}^{\operatorname{joint}}_{\bm{\pi}^{\operatorname{old}}}(\bm{\pi}) =  \sum_{s} \bm{\rho}^{\operatorname{old}}(s) \sum_{\bm{a}} \bm{\pi}(\bm{a} | s) A^{{\on{old}}}(s,\bm{a}), \notag
\end{align}
where $D_{\operatorname{KL}}^{\operatorname{max}}$ denotes the maximum KL-divergence between two policies over states and $\bm{\rho}_{\operatorname{old}}(s)$ is the state distribution under $\bm{\pi}_{\operatorname{old}}$. 
The surrogate function can be optimized to make sure that the original objective improves monotonically. Let $\bm{\pi}^{\on{new}} = \argmax_{\bm{\pi}} S^{\operatorname{TRPO}}(\bm{\pi}, \bm{\pi}^{\operatorname{old}}) $, then we know that $J(\bm{\pi}^{\on{new}}) - J(\bm{\pi}^{\operatorname{old}}) \ge S^{\operatorname{TRPO}}(\bm{\pi}^{\on{new}}, \bm{\pi}^{\operatorname{old}}) \ge S^{\operatorname{TRPO}}(\bm{\pi}^{\on{old}}, \bm{\pi}^{\operatorname{old}}) = 0$, which leads to $J(\bm{\pi}^{\on{new}}) \ge J(\bm{\pi}^{\operatorname{old}})$. So we can define an iteration that $\bm{\pi}^{t+1} = \argmax_{\bm{\pi}} S^{\operatorname{TRPO}}(\bm{\pi}, \bm{\pi}^{t})$, then the sequence $\{J(\bm{\pi}^t)\}$ converges combining with the condition that $J(\bm{\pi})$ is bounded. Following this principle, DPO finds a novel surrogate function that is appropriate for fully decentralized learning:
\begin{align}
    & J({\bm{\pi}})-J({\bm{\pi}^{\operatorname{old}} } ) \ge \sum_{i = 1}^N S^{\on{DPO}}_i(\pi_i,\pi_i^{\on{old}}) \notag \\ 
    & S^{\on{DPO}}_i(\pi_i,\pi_i^{\on{old}}) =  \frac{1}{N} \mathcal{L}^{i}_{\bm{\pi}^{\operatorname{old}}}(\pi_i) -  
          \hat{M}  \sqrt{D^{\max}_{\on{KL}}(\pi^{\on{old}}_{i} \Vert \pi_{i} )} -  C    D^{\max}_{\on{KL}}(\pi^{\on{old}}_{i} \Vert \pi_{i}), \notag  \\
    & \mathcal{L}^{i}_{\bm{\pi}^{\operatorname{old}}}(\pi^i) = \sum_{s} \bm{\rho}^{\operatorname{old}}(s) \sum_{a_i} \pi^i(a_i|s) A_i^{\operatorname{old}}(s,a_i), \notag
\end{align}
where $\hat{M}$ and $C$ are two constants. An important property of $S^{\on{DPO}}_i(\pi_i,\pi_i^{\on{old}})$ is that it can be optimized independently for each agent and make the joint policy improve monotonically, simultaneously. By defining an iteration $\pi_i^{t+1} = \argmax_{\pi_i} S^{\on{DPO}}_i(\pi_i,\pi_i^t)$, the sequence $\{J(\bm{\pi}^t)\}$ improves monotonically. As for the practical algorithm, DPO uses two adaptive coefficients to replace the large constants $\hat{M}$ and $C$ following the practice of PPO \citep{PPO},
\begin{align}
    & \pi^{t+1}_i = \arg \max_{\pi_i} \Big( \frac{1}{N}\mathcal{L}^{i}_{\bm{\pi}^{t}}(\pi^i)- \beta^1_i  \sqrt{D^{\on{avg}}_{\on{KL}}(\pi^{t}_{i} \Vert \pi_{i} )} - \beta^2_i D^{\on{avg}}_{\on{KL}}(\pi^{t}_{i} \Vert \pi_i ) \Big),
    \label{eq:adaptive-obj}
\end{align}
where $\beta^1_i$ and $\beta^2_i$ are the adaptive coefficients, $D^{\on{avg}}_{\on{KL}}$ is the average KL-divergence to replace the maximum KL-divergence $D^{\on{max}}_{\on{KL}}$. 

TVPO \citep{TVPO} solves the non-stationarity problem from the perspective of policy optimization. TVPO is motivated by the difference between the term $\sum_{\bm{a}}\pi_i(a_i|s) \pi^{\on{old}}_{-i}(a_{-i}|s) A^{\on{old}}(s,a_i,a_{-i}) $ in the independent objective $\mathcal{L}^{i}_{\bm{\pi}^{\operatorname{old}}}(\pi^i)$ and the term $\sum_{\bm{a}}\pi_i(a_i|s) \pi_{-i}(a_{-i}|s) A^{\on{old}}(s,a_i,a_{-i}) $ in the joint objective $\mathcal{L}^{\operatorname{joint}}_{\bm{\pi}^{\operatorname{old}}}(\bm{\pi}) $ and proposes novel V-function and Q-function combining with $f$-divergence:
\begin{align}
     & V^{\bm{\pi}}_{\bm{\sigma}}(s) = \frac{1}{N} \sum_i \sum_{a_i}\pi_i(a_i|s) \sum_{a_{-i}} \sigma_{-i}(a_{-i}|s)Q^{\bm{\pi}}_{\bm{\sigma}}(s,a_i,a_{-i}) - \omega  D_f \left( \pi_i(\cdot|s) || \sigma_i(\cdot|s) \right) \notag, \\
     & Q^{\bm{\pi}}_{\bm{\sigma}}(s,a_i,a_{-i}) = r(s,a_i,a_{-i}) + \gamma \mathbb{E}_{s^\prime \sim P(\cdot|s,a_i,a_{-i})} \left[ V^{\bm{\pi}}_{\bm{\sigma}}(s^\prime) \right]. \notag
\end{align}
Given a fixed $\bm{\sigma}$, TVPO defines an iteration as following:
\begin{align} 
     \pi^{\on{new}}_i = \argmax_{\pi_i}\sum_{a_i}\pi_i(a_i|s) \sum_{a_{-i}} \sigma_{-i}(a_{-i}|s)Q^{\bm{\pi}_{\on{old}}}_{\bm{\sigma}}(s,a_i,a_{-i}) - \omega  D_f \left( \pi_i(\cdot|s) \| \sigma_i(\cdot|s) \right),
      \notag
\end{align}
and proves that $V^{\bm{\pi}_{\on{old}}}_{\bm{\sigma}}(s) \le V^{\bm{\pi}_{\on{new}}}_{\bm{\sigma}}$. Based on this result, if we take $\bm{\pi}^{\on{old}} = \bm{\sigma} = \bm{\pi}^t$, $\bm{\pi}^{\on{new}} = \bm{\pi}^{t+1}$, $D_f = D_{\on{TV}}$ ($D_{\on{TV}}$ is the total variation distance) and choose an appropriate $\omega$, then the iteration becomes 
    \begin{align}
        \pi^{t+1}_i &  = \argmax_{\pi_i}\sum_{a_i}\pi_i(a_i|s) \sum_{a_{-i}} \pi^t_{-i}(a_{-i}|s)Q^{\bm{\pi}^t}(s,a_i,a_{-i}) - \omega  D_{\on{TV}} \left( \pi_i(\cdot|s) || \pi^t_i(\cdot|s) \right) \notag \\
        & = \argmax_{\pi_i}\sum_{a_i}\pi_i(a_i|s)Q^{\bm{\pi}^t}_i(s,a_i) - \omega  D_{\on{TV}} \left( \pi_i(\cdot|s) || \pi^t_i(\cdot|s) \right).   \notag
    \end{align}
With this iteration, TVPO further proves that $V^{\bm{\pi}_{t+1} }_{\bm{\pi}_t}(s) \ge V^{\bm{\pi}_{t} }(s) \ge V^{\bm{\pi}_{t} }_{\bm{\pi}_{t-1}}(s) \ge V^{\bm{\pi}_{t-1} }(s) $, which means the sequence $\{V^{\bm{\pi}^t}\}$ improves monotonically and converge to suboptimum. Importantly, this iteration can be executed in a fully decentralized way. In the algorithm of TVPO, there is an issue similar to DPO and TRPO where the large constant $\omega$ may lead to a small stepsize in the gradient update. So TVPO also uses an adaptive coefficient $\beta_i$ to replace $\omega$.

\section{Reward Sum Setting}
\label{sec:sum}

In the reward sum setting, each agent $i$ has an individual reward function $r_i(s,\bm{a},s^\prime)$ and an objective $J_i(\pi_i, \pi_{-i}) = \mathbb{E}_{\bm{\pi}} \left[\sum_t \gamma^t \sum_i r_i(s_t,\bm{a_t},s_{t+1}) \right]$. However, as we consider fully decentralized learning, each agent has no access to the rewards of other agents. Therefore, the reward sum setting degenerates to the \textit{general sum} setting, \textit{i.e.}, $J_i(\pi_i,\pi_{-i}) = \mathbb{E}_{\bm{\pi}} \left[\sum_t \gamma^t r_i(s_t,\bm{a_t},s_{t+1}) \right]$. In the general sum setting, the optimal policies for different agents are usually different and cannot be achieved simultaneously. So the target of the algorithms in the general sum setting is to find the \textbf{Nash equilibrium (NE)}. A joint policy $\bm{\pi}^*$ is a Nash equilibrium if for any agent $i$, $J_i(\pi^*_i, \pi^*_{-i}) = \max_{\pi_i} J_i(\pi_i, \pi^*_{-i}) $. In other words, for any agent $i$, $\pi^*_i$ is the \textbf{best response} of $\pi^*_{-i}$. Unfortunately, previous studies show that the complexity of finding a Nash equilibrium in a general sum game is PPAD-complete \citep{Nash-PPAD}, which means we can hardly find a Nash equilibrium in practice. So the existing studies usually need some assumption about the structure of the game or try to find some weaker equilibrium. In the following, we survey representative and recent studies on the general sum setting and again divide them into value-based methods and policy-based methods.  

\subsection{Value-Based Methods}

In this section, we will introduce two lines of research in the general sum setting: 
We introduce two types of value-based methods: Decentralized Q-learning \citep{DQL} and V-learning \citep{V-learning}. Decentralized Q-learning is proven to asymptotically converge to a Nash equilibrium with a high probability. V-learning is proven to converge to a \textbf{coarse correlated equilibrium (CCE)} with a high probability. CCE is a weaker equilibrium than NE, which allows for the policies to be correlated while NE requires all the policies to be independent. Unlike the asymptotic results of Decentralized Q-learning, V-learning provides the sample complexity analysis for convergence. 

\subsubsection{Decentralized Q-Learning}
Decentralized Q-learning \citep{DQL} relies on the assumption that the general sum game is a \textbf{weakly acyclic game}. If we take all the deterministic policies in a general sum game as the nodes of a graph and there is an edge from the policy $\bm{\pi}^1$ to the policy $\bm{\pi}^2$ if and only if there exists one agent $i$ such that $\pi^2_{-i} = \pi^1_{-i}$ and $\pi^2_i$ is the best response of $\pi^1_{-i}$. A general sum game is a weakly acyclic game if for any policy $\bm{\pi}^0$, there exists a path $(\bm{\pi}^0,\bm{\pi}^1,\cdots,\bm{\pi}^L)$ where $\bm{\pi}^L$ is a Nash equilibrium. 

Given the assumption of the weakly acyclic game, the main idea of Decentralized Q-learning is relatively easy to understand. If an algorithm could satisfy the condition that after $K$ steps, the policy stays at the equilibrium with a probability $p > 0$, then we can repeat the process $M$ times which means the probability of the event that the policy does not stay at the equilibrium is $(1 - p)^M$. Let $M \to \infty$ and we know that the policy will finally converge to a NE with probability $1$. 

Decentralized Q-learning is an algorithm satisfying this condition with the idea of inertia policy update. Inertia policy update means that for any agent $i$, if the policy $\pi_i$ is already the best response of the other agents' policies $\pi_{-i}$, then agent $i$ should keep $\pi_i$ unchanged; otherwise, with probability $\lambda_i$ which corresponds to the inertia in the policy update, agent $i$ will remain  $\pi_i$, and with probability $1 - \lambda_i$, $\pi_i$ will become any other policy uniformly. It is simple to show that the inertia policy update satisfies the condition mentioned above. If the joint policy $\bm{\pi}$ is already an NE, from the inertia policy update we know that the joint policy will not be changed. Otherwise, from the property of the weakly acyclic game, we know that there exists one path $(\bm{\pi}^0=\bm{\pi},\bm{\pi}^1,\cdots,\bm{\pi}^L)$ and the probability $q$ of moving forward one step along the path is positive since there exists at least one situation where one policy becomes the best response in the uniform update and all the other policies remain unchanged from the inertia. So after $L$ steps, the policy reaches the NE $\bm{\pi}^L$ with probability $q^L >0$. Let $p = q^L,K = L$, then we can follow the idea mentioned above to complete the proof. 

The problem remaining for Decentralized Q-learning is to judge whether a joint policy is an NE or a policy $\pi_i$ is the best response of other agents. The solution is Q-learning. Given the policies of other agents $\pi_{-i}$ fixed, if the agent $i$ updates through Q-learning, then we know that the policy $\pi_i$ will converge to the optimal policy which is the best response of $\pi_{-i}$. If all the agents update their policies through Q-learning, the environment becomes non-stationary. So Decentralized Q-learning uses the exploration phase technique, which divides the learning process into several exploration phases. The idea of the exploration phase is similar to on-policy learning.  In each exploration phase, all the policies will be fixed and the samples will only be used to update the Q-function. At the end of each exploration phase, the policy will be updated from the Q-function. It is obvious that if the length of exploration phases $\{t_k\}$ is sufficiently long then the Q-function will be sufficiently accurate and the agents can obtain the correct best response. 

There are several succeeding works for Decentralized Q-learning. Asynchronous Decentralized Q-Learning \citep{ADQL} discusses the situation that each agent has an independent exploration phase sequence $\{t_k^i\}$. Asynchronous Decentralized Q-Learning believes that the shared exploration phase sequence $\{t_k\}$ is a synchronous constraint for decentralized learning and proves that even with independent exploration phase sequence $\{t_k^i\}$, which means the non-stationary problem will arise again, it still has the convergence guarantee. Asynchronous Decentralized Q-Learning requires that $\{t_k^i\}$ satisfies the condition $\exists T,R \in \mathbb{N}, t_k^i \in [T,RT]$, which means that the differences of the update frequency should be limited. Given this condition, the key to the proof of Asynchronous Decentralized Q-Learning is that there exists a sequence of active phase $\{[\tau_k^{\operatorname{min}},\tau_k^{\operatorname{max}}]\}$, which has several properties: (1) all the agents have at least one chance to update in the interval $[\tau_k^{\operatorname{min}},\tau_k^{\operatorname{max}}]$; (2) the total number of all the agent updates is finite and the length of the interval $[\tau_k^{\operatorname{min}},\tau_k^{\operatorname{max}}]$ is finite; (3) all the agents will not update their policies in the interval $(\tau_k^{\operatorname{max}},\tau_{k+1}^{\operatorname{min}})$ and the length of this interval is at least $T/N$. The condition (3) means that if $T$ is sufficiently large, then $(\tau_k^{\operatorname{max}},\tau_{k+1}^{\operatorname{min}})$ is enough for agents to obtain best response from Q-learning. The conditions (1) and (2) mean that the probability of the joint policy moving forward one step along the path in one active phase is positive as there exists one situation where one policy becomes the best response in one update chance and all the policies remain unchanged in other update chance from the inertia.

Independent Team Q-learning \citep{DQL-optimal} tries to find the team optimal policy following the idea of Decentralized Q-learning. A joint policy $\bm{\pi}^*$ is team optimal if for any agent $i$, $J_i(\bm{\pi}^*) = \max_{\bm{\pi}} J_i(\bm{\pi})$. The team optimal policy doesn't always exist and if a general sum game has a team optimal policy then we call it a common interest game.  Independent Team Q-learning finds the team optimal policy in the common interest game through a similar way to inertia policy update. If a joint policy $\bm{\pi}$ is a team optimal policy, with probability $1 - \gamma^i$ the policy $\pi_i$ will stay the same and with probability $\gamma^i$ the policy $\pi_i$ will become any policy uniformly. Otherwise, with probability $1 - \kappa^i$ the policy $\pi_i$ will be changed by a transition kernel $h^i$ which can be chosen by the user freely and with probability $\kappa^i$  will become any policy uniformly. Independent Team Q-learning builds a Markov Chain of the policy and proves that the stationary distribution of this Markov Chain will lie in the set of team optimal policies with probability one if $\gamma^i << \kappa^i$. Independent Team Q-learning evaluates the summation of Q-functions after $k$ updates $S_i^k = \sum_s Q_i^k(s,\pi_i^k(s))$ and uses the condition $S_i^k < \min \{S_i^{k-1}, S_i^{k-2},\cdots,  S_i^{k-W_i}\} + d_i$ to judge whether a joint policy $\bm{\pi}_i^k$ is team optimal, where $W_i$ is a constant and $d_i$ is tolerance of the sub-optimality for agent $i$. This condition is effective when the Q-function is sufficiently accurate and the joint policy has been a team optimal policy in the last $W^i$ updates. The probability of this event is positive as the exploration phase can be sufficiently long and a team optimal policy can be reached by the uniform transition with probability $\gamma^i$ or $\kappa^i$. 

\subsubsection{V-Learning}
V-learning \citep{V-learning} discusses the convergence to CCE in the episodic MDP setting. In the episodic MDP, suppose that the horizon is $H$, then for each time step $h \in \{1,2,\cdots,H\}$, the reward function $r_{i,h}(s,\bm{a},s^\prime)$ and the transition probability $P_h(s^\prime|s,\bm{a})$ are both related to the time step $h$. Moreover, the policy $\pi_{i,h}$ and the value function $V^{\bm{\pi}}_{i,h}$ are also related to the time step $h$. 

The idea of V-learning for finding CCE is to bound the difference between the joint policy $\bm{\pi}$ and its best response $\max_{i} V_{i,1}^{\dag,\pi_{-i}} - V_{i,1}^{\pi_i,\pi_{-i}}$. V-learning uses an optimistic approximation $\bar{V}_{i,h}$ as the upper bound of the best response $V_{i,1}^{\dag,\pi_{-i}}$.  The update rule $\bar{V}_{i,h}$ is 
\begin{align}
    \bar{V}_{i,h}(s_h) \leftarrow (1 - \alpha_t) \bar{V}_{i,h}(s_h) + \alpha_t \left( r_{i,h} +  \bar{V}_{i,h+1}(s_{h+1} + \beta_t)\right) \notag
\end{align}
where $t = N_h(s_h)$ is visitation times of the pair $(h,s_h)$, $\alpha_t$ is the learning rate and $\beta_t > 0$ is the bonus which is key to the optimistic approximation. Moreover, V-learning uses a pessimistic approximation $\underline{V}_{i,h}$ as the lower bound of $V_{i,1}^{\pi_i,\pi_{-i}}$. $\underline{V}_{i,h}$ is similar to $\bar{V}_{i,h}$ but the bonus $ \beta_t$ is replaced with $ -\beta_t$. With carefully designed $\alpha_t$ and $\beta_t$, the upper bound and lower bound are effective and V-learning can use the difference $\bar{V}_{i,h} - \underline{V}_{i,h}$ to control the distance between the current policy and CCE. 

The episodic MDP setting means that the policies $\pi_{i,h}(\cdot|s)$ are relatively independent for each pair $(s,h)$, so the policy update can be executed by the bandit update. V-learning uses the bandit algorithm Follow-the-Regularized-Leader (FTRL) to update the policy $\pi_{i,h}(\cdot|s)$ for each pair $(s,h)$. The bandit update has an important property for V-learning that the regret for the bandit update is bounded and the bound is about the update times $t$. V-learning uses $1 - \frac{r_h + V_h(s)}{H}$ as the loss or reward for the bandit over the pair $(s,h)$ to obtain a proper regret bound for the proof of the convergence.

The algorithm of V-learning is relatively simple. Suppose that training process contains $K$ episodes, then for each episode $k$, the value function $\bar{V}_{i,h}$ is updated by the optimistic bonus and the policy $\pi^k_{i,h}$ is updated by the bandit update. However, the policies $\pi^k_{i,h}$ obtained in the training are not the output policy $\hat{\bm{\pi}}$ that can approximate the CCE and must be saved for calculating the output policy. V-learning has an algorithm for calculating the output policy and the formulation of $\hat{\bm{\pi}}$ is complicated. \citet{V-learning} also prove that the sample complexity of V-learning is $\mathcal{O}(H^5SA_{\operatorname{max}}/\epsilon^2)$ which means that to obtain an output policy within the range of $\epsilon$ from the CCE, V-learning needs $\mathcal{O}(H^5SA_{\operatorname{max}}/\epsilon^2)$ episodes, where $S$ is the number of state and $A_{\operatorname{max}} = \max_i|A_i|$.

There are several related works of V-learning. However, the algorithm and the proof of V-learning are integrated tightly and these related works follow the similar idea of V-learning to obtain the sample complexity, so the changes of these works in the algorithm are relatively small. V-learning OMD \citep{V-learning-OMD} uses mirror descent for the bandit update of the policies instead of FTRL and obtain the sample complexity $\mathcal{O}(H^6SA_{\operatorname{max}}/\epsilon^2)$ which is weaker than V-learning. Stage-based V-learning \citep{stage-based-V-learning} divides the learning process of $V_h(s)$ into several stages for each pair $(s,h)$ according to the visitation times $t = N_h(s)$. $V_h(s)$ will be updated if and only if one stage of $(s,h)$ ends. The length of the stages is a geometric series with a common ratio $1 + 1/H$. Stage-based V-learning can also obtain the sample complexity $\mathcal{O}(H^5SA_{\operatorname{max}}/\epsilon^2)$ as V-learning. \citet{V-learning-FA-1} and \citet{V-learning-FA-2} follow the similar idea of policy replay to extend V-learning from the tabular case to the function approximation case. They both save the learned policies into a buffer and uniformly sample one policy from this buffer for the learning in the new episode. The calculation of the output policy is still needed.

\subsection{Policy-based Method}
The practical algorithms of the policy-based methods are relatively straightforward. Most of them are independent actor-critic or even REINFORCE. So the main contributions of these studies are the discussion and analysis of the convergence to the Nash equilibrium. As we mentioned before, finding a Nash equilibrium in a general sum game is quite difficult so the convergence results of these studies are asymptotic. The main idea behind the proof of the convergence result is to control the difference $\max_{\hat{\pi}_i}J_i(\hat{\pi}_i, \pi_{-i}) - J_i(\pi_i, \pi_{-i})$. The gradient dominance condition is critical for controlling the difference, which means $J_i(\hat{\pi}_i, \pi_{-i}) - J_i(\pi_i, \pi_{-i}) \le M \max_{\pi_i^\prime} \left< \pi_i^\prime -  \pi_i, \nabla_i J_i(\pi_i, \pi_{-i}) 
 \right>$ and $M$ is a constant. Given the gradient dominance condition, the problem of finding a Nash equilibrium can be changed to the problem of controlling the bound $M \max_{\pi_i^\prime} \left< \pi_i^\prime -  \pi_i, \nabla_i J_i(\pi_i, \pi_{-i}) 
 \right>$, \textit{i.e.}, controlling the gradient $\nabla_i J_i(\pi_i, \pi_{-i})$ which can be done by the policy gradient. With the main idea of the policy-based algorithms in the general sum setting, we will introduce some related studies in detail. 

\citet{nash-idv-assumption} discusses the problem of the perspective of occupancy measure $\rho_i$ instead of policy $\pi_i$. The occupancy measure $\rho_i$ and the policy $\pi_i$ can be mutually transformed from each other  by the property $\rho_i(s_i,a_i) = \mu_{\pi_i}(s_i)\pi(a_i|s_i)$ and $\pi(a_i|s_i) = \frac{\rho_i(s_i,a_i)}{\sum_{a_i^\prime}\rho_i(s_i,a_i^\prime)}$, where $\mu_{\pi_i}(s_i)$ is the stationary distribution of the state given the policy $\pi_i$. The benefit of using the occupancy measure $\rho_i$ is that we can rewrite the objective as $J_i(\pi_i, \pi_{-i}) = J_i(\rho_i, \rho_{-i}) = \left< \bm{\rho}, r_i \right> = \left< \rho_i, v_{\rho_{-i}}(r_i) \right>$, where $v_{\rho_{-i}}(r_i) = \mathbb{E}_{\rho_{-i}}\left[r_i(s,a_i,a_{-i})\right]$. It is obvious that $J_i(\rho_i, \rho_{-i})$ is linear over $\rho_i$ and the gradient is $\nabla_i J_i(\rho_i, \rho_{-i}) = v_{\rho_{-i}}(r_i)$, which is a stronger condition than the gradient dominance condition. As for the practical algorithm, \citet{nash-idv-assumption} uses the term $R^k_i(s_i,a_i) = \mathbb{E}\left[ \frac{r_i(s,a_i,a_{-i})}{\pi_i(a_i|s)} \right]$ as the unbiased estimator for the gradient $ v_{\rho_{-i}}(r_i)$. The optimization objective is $\rho^{k+1}_i = \arg\max_{\rho_i} \left< \rho^{k}_i, R^k_i\right> - h_i(\rho_i)$, where $h_i$ is the regularization term. To avoid the zero in the denominator of $\frac{r_i(s,a_i,a_{-i})}{\pi_i(a_i|s)}$, \citet{nash-idv-assumption} limits the occupancy measure $\bm{\rho}$ within the space $P^\delta$, where $\delta > 0$ is constant and  $\bm{\rho} \in P^\delta$ satisfies $\rho_i(s_i,a_i) \ge \delta,\,\forall i \in \{1,2,\cdots,N\}$. We need to point out that the convenient property that $J_i(\rho_i, \rho_{-i})$ is linear over $\rho_i$ is built on the strong assumption about the game structure. \citet{nash-idv-assumption} assumes that the joint state $s$ can be divided into $s=(s_1,s_2,\cdots,s_N)$ and the state transition $P_i(s_i^\prime|s_i,a_i)$ of agent $i$ is independent of other agents' actions and states.

\citet{gradient-play, nips-nash, DPG} all apply policy gradient to find a Nash equilibrium and discuss the theoretical results of policy gradient in the general sum setting. So the practical algorithms of \citet{gradient-play, nips-nash, DPG} are similar. However, these studies provide different theoretical results and we will focus on introducing these contents. \citet{gradient-play} show the first-order stationary policy is an equivalence to the Nash equilibrium, which means the policy $\bm{\pi}^*$ satisfies the condition $\left< \pi_i - \pi_i^*, \nabla_i J(\pi_i^*,\pi_{-i}^*) \right> \le 0,\,\forall i\in \{1,2,\cdots,N\},\,\forall \pi_i \in \Pi_i$. With this property, \citet{gradient-play} proves that if the initial policy $\bm{\pi}^0$ is within a neighborhood of a Nash equilibrium $\bm{\pi}^*$, then the policy sequence $\{\bm{\pi}^t\}$ generated by the policy gradient algorithm will converge to $\bm{\pi}^*$ with high probability. \citet{gradient-play} also provide the analysis of the sample complexity for the local convergence result. \citet{nips-nash} also provide proof of the local convergence result to the first-order stationary policy. Furthermore, \citet{nips-nash} propose the second-order stationary policy which means that a policy $\bm{\pi}^*$ satisfies $(\bm{\pi}^* - \bm{\pi})^T \on{Jac}(\bm{\pi}^*) (\bm{\pi}^* - \bm{\pi}) < 0,\, \forall \bm{\pi} \not = \bm{\pi}^*$, where $\on{Jac}(\bm{\pi}^*)$ is Jacobian of $J$ at $\bm{\pi}^*$. \citet{nips-nash} show that the second-order stationary policy is a sufficient condition for the first-order stationary policy. \citet{nips-nash} also prove a local asymptotic convergence result of the second-order stationary policy which is a stronger theoretical result. \citet{DPG} extend the discussion into the case of the continuous state space and action space. With the property of the equivalence between the first-order stationary policy and the Nash equilibrium, \citet{DPG} change the problem of finding a Nash equilibrium into the problem of variational inequality, which means trying to find a solution $x^*$ satisfying the condition $\left< G(x^*), x^* - x \right>,\,\forall x \in K$, where $K$ is the domain of $x$ and $G(x)$ is a given function. In this problem, $x$ corresponds to the policy $\pi_i$  and the function $G(x)$ corresponds to the gradient $\nabla_i J(\pi_i,\pi_{-i})$. With these preparations, \citet{DPG} design a two-loop algorithm in which the authors sequentially update a constructed strongly monotone variational inequality in the outer loop by updating a proximal parameter and employ a single-call extra-gradient algorithm in the inner loop for solving the constructed variational inequality. Moreover, \citet{DPG} provide global asymptotic convergence results of the Nash equilibrium which means the initial policy is not required to stay within the neighborhood of a Nash equilibrium. Instead,  the results in \citet{DPG} need the assumption that the policy space is a nonempty compact convex which is a much looser condition than the neighborhood condition.

\section{Discussion}
\label{sec:discuss}

\subsection{Open Questions}

The research on fully decentralized MARL is still preliminary. There are many open questions worth exploring:

\begin{itemize}

\item \textbf{Optimal joint policy}. The existing works in the shared reward setting mainly focus on the convergence result. Value-based methods can converge to optimal joint policy in both deterministic and stochastic environments, \textit{i.e.}, BQL, while policy-based methods can only guarantee the convergence to suboptimal joint policy, \textit{e.g.}, TVPO. Therefore, how to devise a policy-based algorithm that has the convergence to optimal joint policy is still an open question.  

\item \textbf{Sample complexity}. Most convergence results in the shared reward setting and the general sum setting are asymptotic. So providing the analysis of the sample complexity for fully decentralized algorithms may be an interesting direction for future work. Moreover, some decentralized algorithms are still on-policy which may be troubled with the poor sample efficiency, especially in the MARL setting. Proposing novel algorithms with better sample efficiency can be a critical open question. 

\item \textbf{Coordination}. Most analysis of optimal joint policy is based on the assumption that there is only one optimal joint policy. When there are multiple optimal joint actions at some state, if each agent arbitrarily selects one of the optimal independent actions, the joint action might not be optimal. It is hard to learn a coordinated policy in a fully decentralized way. Existing coordination methods require information exchange between agents \citep{zhang2013coordinating,bohmer2020deep,li2021deep}. Decentralized coordination without any communication or pre-defined rules is quite a challenge.

\item \textbf{Offline decentralized MARL}. In offline decentralized MARL, the agents cannot interact with the environment to collect experiences but have to learn from offline datasets pre-collected by behavior policies. Unlike single-agent offline RL, where the main cause of value estimation error is out-of-distribution actions, the agents also suffer from the bias between offline transition probabilities and online transition probabilities. Still, the dataset of agent $i$ does not contain the actions of other agents and the agents cannot share information during train and execution. Therefore, from the perspective of agent $i$, the offline transition probability in the dataset is:
$$P_{\mathcal{B}_i}\left(s^{\prime} | s, a_i\right)={\sum}_{{a}_{-i}}P\left(s^{\prime} | s, a_i,{a}_{-i}\right) {\pi}_{\mathcal{B}_{-i}}({a}_{-i} | s),$$
which depends on other agents' behavior policies ${\pi}_{\mathcal{B}_{-i}}$. The online transition probability during execution is:
$$P_{\mathcal{E}_i}\left(s^{\prime} | s, a_i\right)={\sum}_{{a}_{-i}}P\left(s^{\prime} | s, a_i,{a}_{-i}\right) {\pi}_{\mathcal{E}_{-i}}({a}_{-i} | s),$$
which depends on other agents' learned policies ${\pi}_{\mathcal{E}_{-i}}$. As the learned policies may greatly deviate from behavior policies, there is a large bias between offline and online transition probabilities, which leads to value estimation error. MABCQ \citep{jiang2021offline} tries to reduce the bias by normalizing the offline transition probabilities and increasing the transition probabilities of high-value states. OTC \citep{jiang2023online} corrects the offline transition probabilities using limited online experiences. More theoretical analysis and practical methods are expected in this direction.  

\end{itemize}

\subsection{Partial Observation}

The incomplete information about the state caused by the partial observation hinders the process of finding the optimal policy in POMDP. The computation complexity of POMDP has been studied for decades. \citet{1987-POMDP} show that solving the POMDP with the model information is a PSPACE-complete problem which means it is less likely to be solved within polynomial time than the NP-complete problem. \citet{2000-POMDP} take one step further to show that finding the optimal policy in the POMDP is a $\on{NP}^{\on{PP}}$-complete problem, where $\on{NP}^{\on{PP}}$ is a class of complexity between NP and PSPACE. \citet{2011-POMDP} show solving the POMDP is an NP-hard problem. On the other hand, in fully decentralized learning, researchers focus on the convergence of algorithms facing the challenge of the non-stationary problem, either in the shared reward setting or in the general sum setting. So combining the theoretical analysis with the partial observation can be notoriously difficult and the existing works all provide the proof of convergence from the perspective of the state instead of the observation. Partial observation is an important property of POMDP and is worth more attention from the MARL community. But we also would like to appeal to the community to be more tolerant of the progress in fully decentralized learning.  

\subsection{CTDE vs. Fully Decentralized Learning}

Why is fully decentralized learning necessary as we already have CTDE? This is the most commonly asked question. Here we discuss the situations where CTDE is preferred and the situations where decentralized learning should be considered. When the cooperation task is fixed and centralized modules are allowed, CTDE methods can achieve stronger performance since they guarantee convergence to the optimal joint policy, do not suffer from non-stationarity, and have better sample complexity. However, there are some cases where the information of all agents is not available due to network or privacy. Taking autonomous vehicles as an example, agents might belong to different companies and cannot share action information. Therefore, we can only use decentralized learning. Moreover, it is challenging for CTDE methods to handle the varying agent numbers and unknown policies of other agents in open-ended environments, \textit{e.g.}, autonomous vehicles, robots, and online games. Without the constraints of centralized modules, decentralized learning has a high potential in open-ended environments.

\subsection{Unified Reinforcement Learning}

When there is only one agent in the environment, the cooperative MARL setting will degenerate into a single-agent RL setting. Naturally, we expect that the decentralized MARL algorithms can still guarantee convergence to optimal policy when applied in single-agent tasks. This can provide us with a unified perspective of both single-agent RL and multi-agent RL. For example, in single-agent environments, there is only one possible transition distribution in BQL, so BQL degenerates into vanilla Q-learning. In open-ended environments, an agent may cooperate with other agents in certain states while being able to complete sub-tasks independently in other states. A unified reinforcement learning framework allows each agent to be trained by the same algorithm, eliminating the need to switch algorithms based on different scenarios, so it is suitable for learning in open-ended environments. However, the theory behind this unified perspective and the associated practical algorithms require further comprehensive investigation.


\bibliography{main}
\bibliographystyle{tmlr}


\end{document}

%% file: math_commands.tex
\usepackage{amsmath,amsfonts,bm}

\def\eqref#1{equation~\ref{#1}}

\def\1{\bm{1}}

\DeclareMathAlphabet{\mathsfit}{\encodingdefault}{\sfdefault}{m}{sl}
\SetMathAlphabet{\mathsfit}{bold}{\encodingdefault}{\sfdefault}{bx}{n}

\DeclareMathOperator*{\argmax}{arg\,max}